# CAFA-evaluator: A Python Tool for Benchmarking Ontological Classification Methods

Damiano Piovesan[1,*], Davide Zago[1], Parnal Joshi[2,3], M. Clara De Paolis Kaluza[4], , Mahta Mehdiabadi[1], Rashika Ramola[4], Alexander Miguel Monzon[5], Walter Reade[6], Iddo Friedberg[3], Predrag Radivojac[4], Silvio C. E. Tosatto[1]

[1]Department of Biomedical Sciences, University of Padova, Padova 35121, Italy.
[2]Program in Bioinformatics and Computational Biology, Iowa State University, Ames IA, 50011 USA.
[3]Department of Veterinary Microbiology and Preventive Medicine,  Iowa State University, Ames IA, 50011 USA.
[4]Khoury College of Computer Sciences, Northeastern University, Boston, MA 02115, USA.
[5]Department of Information Engineering, University of Padova, Padova 35121, Italy.
[6]Kaggle, San Francisco, CA, USA

*To whom correspondence should be addressed.

## Abstract
We present CAFA-evaluator, a powerful Python program designed to evaluate the performance of prediction methods on targets with hierarchical concept dependencies. It generalizes multi-label evaluation to modern ontologies where the prediction targets are drawn from a directed acyclic graph and achieves high efficiency by leveraging matrix computation and topological sorting. The program requirements include a small number of standard Python libraries, making CAFA-evaluator easy to maintain. The code replicates the Critical Assessment of protein Function Annotation (CAFA) benchmarking, which evaluates predictions of the consistent subgraphs in Gene Ontology. Owing to its reliability and accuracy, the organizers have selected CAFA-evaluator as the official CAFA evaluation software.

**Availability and implementation**: https://pypi.org/project/cafaeval
**Contact**: damiano.piovesan@unipd.it

## Introduction

Translating experimental data into biological knowledge remains a slow process despite the rapid accumulation of data in modern biology. Manually curated databases are the primary source of such knowledge due to their thorough standardization of integrated information, often organized into ontological annotations (The Gene Ontology Consortium, 2019). The automated prediction of ontological annotations has become widely adopted in knowledge bases. As a result, ensuring a reliable evaluation of the predicted information remains crucial.

The Critical Assessment of protein Function Annotation (CAFA) initiative provides a well-defined framework for managing hierarchical data and independently evaluates Gene Ontology (GO) prediction methods (Radivojac et al., 2013; Jiang et al., 2016; Zhou et al., 2019). Since its first edition, the CAFA experiment has stimulated a number of theoretical studies about GO prediction and its evaluation (Clark and Radivojac, 2013; Peng et al., 2018).

Despite the significant impact of CAFA, the development of novel function prediction methods suffers from the lack of an easy-to-use tool for internal benchmarking. Existing solutions are problematic due to missing documentation, hampering their maintenance, portability, development, and use by the scientific community. Moreover, these solutions are tailored specifically for GO terms and the CAFA challenge, incorporating numerous hard-coded parameters.

The CAFA-evaluator package addresses these issues by being easy to use and maintain, fully documented, fast, and generic. It can be used with any type of ontology and annotation, and the dataset processing is entirely separated from the evaluation stage. Additionally, the input format is straightforward. The software has been tested against CAFA2 and CAFA3 data, replicating the exact results provided in their corresponding publications (Jiang et al., 2016; Zhou et al., 2019). CAFA-evaluator has been recently adopted as the official evaluation tool for the CAFA5 challenge hosted on Kaggle.

The CAFA-evaluator software is open source and freely available for download from GitHub and PyPI. The GitHub repository also includes a detailed Wiki section that offers a comprehensive explanation of the algorithm. This Wiki provides valuable insights into the software and offers concrete examples that demonstrate the impact of selecting different parameters during the final evaluation. It serves as a valuable resource for understanding the software and its functionality.

## Implementation

The CAFA-evaluator repository includes a Python library and a user-friendly command-line interface for generating all evaluations and a Python notebook for plotting the results. The evaluation module calculates the F-measure, weighted F-measure, and semantic distance (S-score), as well as precision-recall and remaining uncertainty-misinformation curves, as described in (Jiang et al., 2016). The package requires only three standard Python libraries: Numpy, Pandas, and Matplotlib, with the latter being necessary only for generating plots.

**Input and calculation**

The CAFA-evaluator workflow is shown in Figure 1. The software requires three inputs: an ontology OBO file, a ground truth file, and the path to the folder containing the prediction file(s). Optionally, it also accepts an information accretion file, which triggers the generation of weighted measures such as weighted precision, recall, F-measure, and S-score.

All input files undergo internal parsing, and predictions are filtered to include only those targets present in the ground truth and those terms that are part of the input ontology. When terms are associated with a "namespace", also called "aspect" or "sub-ontology", different namespaces are treated as independent ontologies, and both the ground truth and predictions are split accordingly. Namespaces with multiple roots are managed without problems and it is possible to exclude root terms from the evaluation.

The algorithm stores three sparse matrices in memory: the ontology graph as an adjacency matrix, a boolean n x m matrix, where n is the number of targets and m is the number of ontology terms, representing the ground truth, and a matrix of the same size (or smaller if some targets are missing) including the prediction scores. Multiple prediction files, each corresponding to a different method, are processed one by one to release the memory associated with the third matrix.

Both the predictions and the ground truth annotations are always propagated up to the ontology root(s). By default, however, prediction scores are propagated without overwriting parents' scores, as in CAFA. Optionally, the maximum score over all direct children terms can be propagated to their common parent term. The ontology graph is topologically sorted at the parsing time, allowing the propagation to be calculated in linear time, solely depending on the size of the ontology, which is always the same for all prediction files and is loaded in memory at the beginning.

Confusion matrices are calculated per target and per threshold, i.e. separately by considering predicted terms with a score above the threshold. By default, one hundred evenly spaced cutoffs in the range [0-1) are considered, but more cutoffs can be set by the user; e.g. to capture all unique score predictions for a method. Calculation time depends on the number of threshold cutoffs. The software is parallelized so that blocks of thresholds can be calculated in different threads.

The tool incorporates both macro- and micro-averaging techniques. The macro-averaging approach follows the traditional CAFA method, where metrics are calculated individually for each target (confusion matrix) and then averaged across all targets. Conversely, the micro-averaging approach involves averaging the confusion matrices over the number of targets before calculating the metrics. These two approaches provide different perspectives on the evaluation process and offer a comprehensive analysis of the software's performance.

Additionally, the user can decide whether to normalize considering all ground truth targets, i.e. penalizing methods with low coverage, or considering only the predicted targets. By default, the program normalizes the recall by the number of ground truth targets and the precision by the number of predicted targets, as in CAFA.

When the information accretion file is provided, the confusion matrix is calculated after the terms are weighted by their information accretion. This approach avoids returning the simple count as in the confusion matrix when calculating the graph intersection. Other options control the inclusion or exclusion of root (orphan) terms from the evaluation and limit the number of processed terms per protein and namespace. The latter is particularly useful when prediction methods include a large number of predicted terms per target and when the number of targets is large. In any case, the number of considered terms does not affect the computation or memory usage. More information about the impact of the parameters, a detailed workflow of the algorithm along with explanatory examples are provided in the Wiki of the CAFA-evaluator GitHub repository.

**Output**

The CAFA-evaluator software generates multiple output objects, including a table with an evaluation row for each method, namespace, and threshold. It also generates an object for F-measure, S-score, and weighted F-measure, reporting the rows with the corresponding best performance. The software also includes a function to store the output into tabular files. Finally, it streams basic execution information, such as timestamps and statistics about the number of processed targets and terms.

The evaluation output table can be used as input for the Python notebook to generate curve plots. The notebook accepts an optional file with the name of the team associated with each prediction file. When this information is provided, only one prediction per team and ontology is selected, as in CAFA. Additionally, prediction files can be associated with a different name, which will be displayed in the plots.

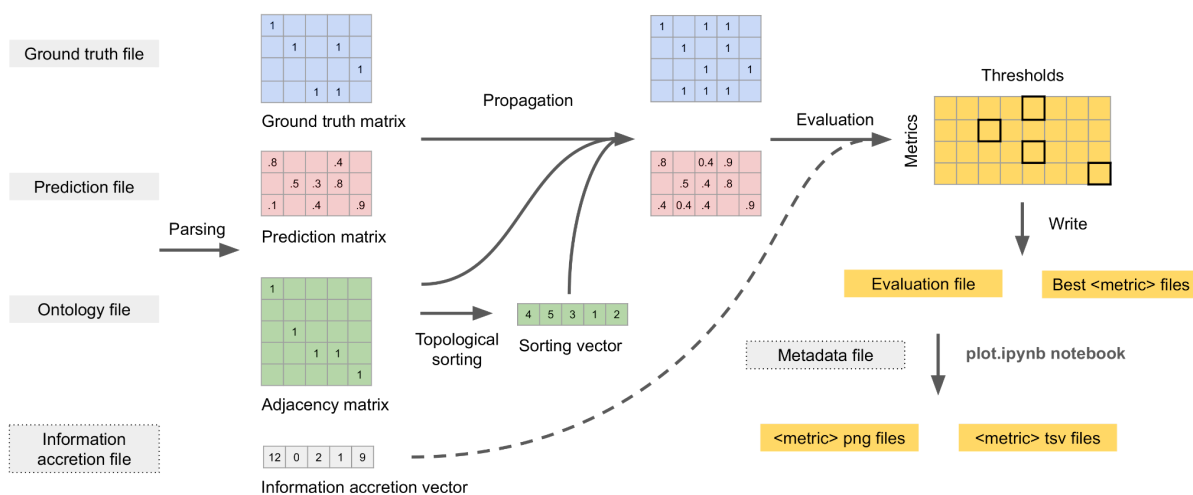

**Figure 1. CAFA-evaluator workflow.**
Gray boxes represent input files, while yellow boxes represent output files. Optional input files are outlined with a dashed line. Internal data structures are depicted as matrices and vectors, with example values provided. Arrows indicate logical processes, often corresponding to code functions. At the end of the workflow, image files are generated using a Jupyter Notebook (plot.ipynb), which takes the output of the CAFA-evaluator library as input.

## Summary
The CAFA-evaluator software is an easy-to-use, generic, and well-documented tool designed for benchmarking function prediction methods using any type of ontology and annotation. It requires an ontology OBO file, a ground truth file, a prediction file, and can optionally accept an information accretion file. The software uses internal parsing to filter predictions and generate multiple output files, including a table with an evaluation row for each method, namespace, and threshold, as well as separate files for F-measure, S-score, and weighted F-measure. The software has been tested against CAFA2 and CAFA3 data and has been adopted as the official evaluation tool for the CAFA5 challenge hosted on Kaggle.

**Funding**
ELIXIR, the research infrastructure for life-science data. This publication is partially based upon work from COST Action ML4NGP (CA21160), supported by COST (European Cooperation in Science and Technology). Funded by the European Union through NextGenerationEU, PNRR project ELIXIRxNextGenIT (IR0000010, CN00000041). Italian Ministry of Education and Research through the NextGenerationEU fund PRIN 2022 project: PLANS (2022W93FTW). Funding for open access charge: University of Padova.

Conflict of Interest: none declared.

## References

Clark,W.T. and Radivojac,P. (2013) Information-theoretic evaluation of predicted ontological annotations. Bioinforma. Oxf. Engl., 29, i53–i61.

The Gene Ontology Consortium (2019) The Gene Ontology Resource: 20 years and still GOing strong. Nucleic Acids Res., 47, D330–D338.

Jiang,Y. et al. (2016) An expanded evaluation of protein function prediction methods shows an improvement in accuracy. Genome Biol., 17, 184.

Peng,Y. et al. (2018) Enumerating consistent sub-graphs of directed acyclic graphs: an insight into biomedical ontologies. Bioinforma. Oxf. Engl., 34, i313–i322.

Radivojac,P. et al. (2013) A large-scale evaluation of computational protein function prediction. Nat. Methods, 10, 221–227.

Zhou,N. et al. (2019) The CAFA challenge reports improved protein function prediction and new functional annotations for hundreds of genes through experimental screens. Genome Biol., 20, 244.